\documentclass[aps,prl,twocolumn,showpacs,floatfix,superscriptaddress,nobalancelastpage,footinbib]{revtex4-1}
\usepackage{graphicx}
\usepackage{multirow}
\usepackage{amssymb}
\usepackage{amsmath}
\usepackage{color}
\usepackage{url}
\usepackage{epstopdf}

\usepackage[urlcolor=blue, hyperindex, colorlinks, bookmarks=true]{hyperref}
\hypersetup{
  linkcolor=red,
  citecolor=blue
}

\newcommand{\red}{\color[rgb]{0.8,0,0}}

\newcommand\scalemath[2]{\scalebox{#1}{\mbox{\ensuremath{\displaystyle #2}}}}

\usepackage{microtype}

\begin{document}

\title{Cross-resonance interactions between superconducting qubits with variable detuning}

\author{Matthew Ware}
\email[E-mail: ]{meware@syr.edu}
\thanks{Current address: Raytheon BBN Technologies, Cambridge, MA 02138, USA}
\affiliation{Department of Physics, Syracuse University, Syracuse, NY 13244-1130, USA}

\author{Blake R. Johnson}
\thanks{Current address: Rigetti Computing, 775 Heinz Ave., Berkeley, CA 94710, USA}
\affiliation{Raytheon BBN Technologies, Cambridge, MA 02138, USA}

\author{Jay M. Gambetta}
\affiliation{IBM T.J. Watson Research Center, Yorktown Heights, NY 10598, USA}

\author{Thomas A. Ohki}
\affiliation{Raytheon BBN Technologies, Cambridge, MA 02138, USA}

\author{Jerry M. Chow}
\affiliation{IBM T.J. Watson Research Center, Yorktown Heights, NY 10598, USA}

\author{B. L. T. Plourde}
\affiliation{Department of Physics, Syracuse University, Syracuse, NY 13244-1130, USA}

\date{\today}


\begin{abstract} Cross-resonance interactions are a promising way to implement all-microwave
two-qubit gates with fixed-frequency qubits. In this work, we study the dependence of the
cross-resonance interaction rate on qubit-qubit detuning and compare with a model that
includes the higher levels of a transmon system. To carry out this study we employ two
transmon qubits---one fixed frequency and the other flux tunable---to allow us to vary the
detuning between qubits. We find that the interaction closely follows a three-level model of the transmon,
thus confirming the presence of an optimal regime for cross-resonance gates.
\end{abstract}

\maketitle

\section{I. Introduction}
\setcounter{secnumdepth}{1}

Superconducting qubits are a promising experimental approach to scalable quantum information
processing \cite{Clarke2008, Devoret2013}, with recent advances \cite{Chow2014, Barends2014,
Corcoles2015} demonstrating fidelity of control of superconducting qubit systems near
the fault tolerance threshold for surface code error correction \cite{Fowler2012}. While
single-qubit gate fidelity is already at or below fault tolerance thresholds for many error
correction schemes, two-qubit gate performance, necessary for universal quantum computing,
has proven more difficult to improve. The cross resonance (CR) effect \cite{Rigetti2010,
Paraoanu2006} is a promising resource for a high-fidelity two-qubit gate because it is
compatible with fixed-frequency qubits, thus eliminating decoherence effects from flux
noise. The CR effect appears when two qubits have a fixed coupling and one qubit, the
control qubit, is driven at the frequency of the second qubit, the target qubit. During
driven evolution, the target qubit undergoes Rabi oscillations between its two lowest-energy
eigenstates at a rate dependent on the state of the control qubit. This effect can be
used as a primitive for a two-qubit controlled-NOT gate between the two qubits. 
This allows for a non-local all-microwave entangling gate \cite{Chow2011} with high
fidelity \cite{Corcoles2013} between fixed-frequency qubits with fixed coupling
\cite{Majer2007}.

In order to further optimize CR gates, we study the dependence of the CR rate
on qubit-qubit detuning, comparing it to a model including energy states outside 
the two-state manifold. To accomplish this, we use a fixed-frequency transmon as the control
qubit and a two-junction flux-tunable transmon as the target, biasing the target qubit at
frequencies both above and below the control qubit's transition frequency. By measuring the
CR rate versus drive strength at different detunings, we observe regimes where the CR rate
is enhanced due to the presence of higher energy levels in the transmon, as well as regimes
where the CR rate is reduced. A complete description of the cross-resonance effect must
include the effects of these higher qubit levels. For the purposes of this study, we focus
on the speed of CR-induced rotations and not other effects, such as leakage outside the
qubit manifold, which would also participate in the fidelity of a two-qubit gate.

The paper is organized as follows. In Sec.~\ref{sec:theory} we discuss the theory of cross-resonance
interactions between two transmon qubits accounting for the role of higher levels. Our device
design and experimental setup is described in Sec.~\ref{sec:exp}. The details of the measurement are
outlined in Sec.~\ref{sec:meas}. The data analysis and processing are discussed in Sec.~\ref{sec:analysis}.

\section{II. Theory of cross-resonance with higher levels}
\label{sec:theory}
For two driven qubits coupled to a common cavity in the lab frame, the Hamiltonian is
\begin{equation} \begin{split} H/\hbar = \frac{1}{2}\omega_1 \hat{Z}\hat{I} + \Omega_1{\rm
cos}(\omega^{{\rm rf}}_{1}t + \phi_1)\hat{X}\hat{I} + \\ \frac{1}{2}\omega_2 \hat{I}\hat{Z} +
\Omega_2{\rm cos}(\omega^{{\rm rf}}_{2}t + \phi_2)\hat{I}\hat{X} + \frac{1}{2} J
\hat{X}\hat{X}, \label{2qHam} \end{split} \end{equation}
where $\{ X,Y,Z,I \}^{\otimes 2}$ are qubit Pauli operators ordered by qubit subspace and $J$ is the residual qubit-qubit coupling.
This Hamiltonian can be diagonalized to yield two new qubits with shifted frequencies
$\omega_1 + J/\Delta$ and $\omega_2 - J/\Delta$ with $\Delta = \omega_1-\omega_2$. A drive at Q2 with a frequency
$\omega_1 + J/\Delta$ will drive Rabi oscillations in Q1 at a rate dependant on the
state of Q2.  This is the basic cross-rsonance effect as described in~\cite{Rigetti2010}.  To model the cross resonance effect including higher energy levels of the transmon we
consider the Duffing Hamiltonian for the case above:
   \begin{equation}\begin{split}
   \hat H_\mathrm{sys} =& \hbar\left[\tilde\omega_1 \hat{b}^\dagger \hat{b}+\frac{1}{2}\delta_1{\hat{b}^\dagger \hat{b}}(\hat{b}^\dagger \hat{b}-1)\right]\\&+  \hbar\left[\tilde\omega_2\hat{c}^\dagger \hat{c}+\frac{1}{2}\delta_2{\hat{c}^\dagger \hat{c}}(\hat{c}^\dagger \hat{c}-1)\right]\\&+ \hbar J(\hat b\hat c^\dagger + \hat b^\dagger \hat c)
   \label{eq_H_dispersive3}\end{split}
   \end{equation} where $\omega_i$ are the resonance frequencies for the 0-1 transition of the transmon $i$, $\delta_i$ is the anharmonicity of transmon $i$, $J$ is the effective coupling rate, and $\hat b$ ($\hat c$) are the annihilation operators for transmon 1 and 2 respectively. Control of the qubits are modeled by
    \begin{equation}\begin{split}
      \hat H_\mathrm{cont} =&\mathcal{E}_1 (\hat b + \hat b^\dagger)+ \mathcal{E}_2 (\hat c + \hat c^\dagger).\\
      \label{eq_H_dispersive3}\end{split}
      \end{equation}
    For more information on how this can be derived from two qubits inside a cavity see Refs.~\cite{Gambetta2013,Magesan2019}. In a physical systems we
usually measure in the energy eigenstates and as such it is much simpler to defined this basis as the computation basis. Treating $J$ as a perturbation
to second order and truncating the system at 3 excitations we find
     \begin{equation}\begin{split}
       \hat H_\mathrm{sys} =&\left[\omega_2-\frac{J^2}{\Delta }\right]|01\rangle \langle 01|+ \left[\omega_1+\frac{J^2}{\Delta }\right]|10\rangle \langle 10|
       \\&+\left[\omega_1+\omega_2+\zeta\right] |11\rangle \langle 11|\\&+ \left[2\omega_2 +\delta_2+\frac{2J^2}{\delta_2-\Delta }\right]|02\rangle \langle 02|\\&+
       \left[2\omega_1 +\delta_1+\frac{2J^2}{\delta_1+\Delta}\right]|20\rangle \langle 20| \\&+\left[3\omega_2 +3\delta_2+\frac{3J^2}{2\delta_2-\Delta }\right]|03\rangle \langle 03|\\&+\left[2\omega_2 +\delta_2+\omega_1+
              \frac{J^2 (\Delta-3\delta_1-5\delta_2)}{(2\delta_2-\Delta)(\Delta +\delta_1-\delta_2)}
              \right]|12\rangle \langle 12|
           \\&+\left[2\omega_1 +\delta_1+\omega_2+
                        \frac{J^2 (\Delta+5\delta_1+3\delta_2)}
                        {(2\delta_1+\Delta)(\Delta +\delta_1-\delta_2)}
                        \right]|21\rangle \langle 21|
    \\&+
                  \left[3\omega_1 +3\delta_1+\frac{3J^2}{2\delta_1+\Delta}\right]|30\rangle \langle 30|
               \end{split}
    \end{equation} where $\zeta=\frac{2 J^2 (\delta_1+\delta_2)}{(\Delta +\delta_1) (\Delta -\delta_2)}$
   and more importantly the control Hamiltonians becomes
   \begin{widetext}
   \begin{equation*}
   H_1 =\left(
   \scalemath{0.75}{
   \begin{array}{cccccccccc}
    0 & -\frac{J}{\Delta } & 1 & 0 & 0 & 0 & 0 & 0 & 0 & 0 \\
    -\frac{J}{\Delta } & 0 & 0 & 1 & \frac{\sqrt{2} J}{\delta_2 -\Delta } & -\frac{\sqrt{2} J \delta_1 }{\Delta  (\delta_1 +\Delta )} & 0 & 0 & 0 & 0 \\
    1 & 0 & 0 & J \left(\frac{1}{\Delta }-\frac{2}{\delta_1 +\Delta }\right) & 0 & \sqrt{2} & 0 & 0 & 0 & 0 \\
    0 & 1 & J \left(\frac{1}{\Delta }-\frac{2}{\delta_1 +\Delta }\right) & 0 & 0 & 0 & 0 & -\frac{\sqrt{2} J (\delta_2 +\delta_1 -\Delta )}{(\delta_2 -\Delta ) (-\delta_2 +\delta_1 +\Delta )} & \sqrt{2} & -\frac{\sqrt{6} J \delta_1 }{(\delta_1 +\Delta ) (2 \delta_1 +\Delta )} \\
    0 & \frac{\sqrt{2} J}{\delta_2 -\Delta } & 0 & 0 & 0 & 0 & \frac{\sqrt{3} J}{2 \delta_2 -\Delta } & 1 & \frac{2 J \delta_1 }{(\delta_2 -\Delta ) (-\delta_2 +\delta_1 +\Delta )} & 0 \\
    0 & -\frac{\sqrt{2} J \delta_1 }{\Delta  (\delta_1 +\Delta )} & \sqrt{2} & 0 & 0 & 0 & 0 & 0 & \frac{J (\delta_1 -\Delta )}{(\delta_1 +\Delta ) (2 \delta_1 +\Delta )} & \sqrt{3} \\
    0 & 0 & 0 & 0 & \frac{\sqrt{3} J}{2 \delta_2 -\Delta } & 0 & 0 & 0 & 0 & 0 \\
    0 & 0 & 0 & -\frac{\sqrt{2} J (\delta_2 +\delta_1 -\Delta )}{(\delta_2 -\Delta ) (-\delta_2 +\delta_1 +\Delta )} & 1 & 0 & 0 & 0 & 0 & 0 \\
    0 & 0 & 0 & \sqrt{2} & \frac{2 J \delta_1 }{(\delta_2 -\Delta ) (-\delta_2 +\delta_1 +\Delta )} & \frac{J (\delta_1 -\Delta )}{(\delta_1 +\Delta ) (2 \delta_1 +\Delta )} & 0 & 0 & 0 & 0 \\
    0 & 0 & 0 & -\frac{\sqrt{6} J \delta_1 }{(\delta_1 +\Delta ) (2 \delta_1 +\Delta )} & 0 & \sqrt{3} & 0 & 0 & 0 & 0 \\
   \end{array}
   }
   \right)
 \end{equation*}
        and
        \begin{equation*}
       H_2= \left(
       \scalemath{0.82}{
        \begin{array}{cccccccccc}
         0 & 1 & \frac{J}{\Delta } & 0 & 0 & 0 & 0 & 0 & 0 & 0 \\
         1 & 0 & 0 & \frac{J (\delta_2 +\Delta )}{\Delta  (\Delta -\delta_2 )} & \sqrt{2} & 0 & 0 & 0 & 0 & 0 \\
         \frac{J}{\Delta } & 0 & 0 & 1 & \frac{\sqrt{2} J \delta_2 }{(\delta_2 -\Delta ) \Delta } & \frac{\sqrt{2} J}{\delta_1 +\Delta } & 0 & 0 & 0 & 0 \\
         0 & \frac{J (\delta_2 +\Delta )}{\Delta  (\Delta -\delta_2 )} & 1 & 0 & 0 & 0 & -\frac{\sqrt{6} J \delta_2 }{2 \delta_2 ^2-3 \Delta  \delta_2 +\Delta ^2} & \sqrt{2} & \frac{\sqrt{2} J (\delta_2 +\delta_1 +\Delta )}{(\delta_1 +\Delta ) (-\delta_2 +\delta_1 +\Delta )} & 0 \\
         0 & \sqrt{2} & \frac{\sqrt{2} J \delta_2 }{(\delta_2 -\Delta ) \Delta } & 0 & 0 & 0 & \sqrt{3} & \frac{J (\delta_2 +\Delta )}{2 \delta_2 ^2-3 \Delta  \delta_2 +\Delta ^2} & 0 & 0 \\
         0 & 0 & \frac{\sqrt{2} J}{\delta_1 +\Delta } & 0 & 0 & 0 & 0 & -\frac{2 J \delta_2 }{(\delta_1 +\Delta ) (-\delta_2 +\delta_1 +\Delta )} & 1 & \frac{\sqrt{3} J}{2 \delta_1 +\Delta } \\
         0 & 0 & 0 & -\frac{\sqrt{6} J \delta_2 }{2 \delta_2 ^2-3 \Delta  \delta_2 +\Delta ^2} & \sqrt{3} & 0 & 0 & 0 & 0 & 0 \\
         0 & 0 & 0 & \sqrt{2} & \frac{J (\delta_2 +\Delta )}{2 \delta_2 ^2-3 \Delta  \delta_2 +\Delta ^2} & -\frac{2 J \delta_2 }{(\delta_1 +\Delta ) (-\delta_2 +\delta_1 +\Delta )} & 0 & 0 & 0 & 0 \\
         0 & 0 & 0 & \frac{\sqrt{2} J (\delta_2 +\delta_1 +\Delta )}{(\delta_1 +\Delta ) (-\delta_2 +\delta_1 +\Delta )} & 0 & 1 & 0 & 0 & 0 & 0 \\
         0 & 0 & 0 & 0 & 0 & \frac{\sqrt{3} J}{2 \delta_1 +\Delta } & 0 & 0 & 0 & 0 \\
        \end{array}
        }
        \right).\end{equation*}
   \end{widetext}
Here we clearly see the cross resonance effect. Looking at the top 4 by 4 block of $H_1$ there are now matrix elements that drive both qubit 1 and qubit 2
with the elements on qubit 2 being different depending on the state of qubit 1: $-J/\Delta$ and $J(\delta_1-\Delta) /\Delta(\delta_1+\Delta)$. The limit $\delta_1\rightarrow\infty$ gives completely the opposite sign resulting in both states rotating in opposite directions while as
$\delta_1\rightarrow 0 $ both states rotate in the same direction giving no conditional operation~\cite{deGroot2012,Rigetti2010,Chow2011}.

During the Rabi drive, the effective Hamiltonian becomes
\begin{equation}\begin{split}
      \hat H^{\mathrm{D_2}}_{\mathrm{dr_1}} = \hbar\epsilon(t)[\hat Z \hat I - \nu_1
      \hat I \hat X - \mu_1 \hat Z \hat X]
      \label{H_eff}\end{split}
\end{equation}
where $\epsilon$ is the drive amplitude. The $ZI$ term represents the ac-Stark
shift from driving the target qubit, Q1, off-resonant with its 0-1 transition frequency.
The second term takes into account the classical and quantum crosstalk that inevitably induces
rotations on the target qubit independent of the control qubit state. As noted elsewhere \cite{Corcoles2013}, these
interactions do not degrade the conditional $ZX$ term as these terms commute. We have intentionally omitted a $ZZ$ crosstalk term to focus on the direct drive terms~\cite{Gambetta2013,Patterson2019}.  The $ZX$ prefactor $\mu$ is recovered by looking at the $H_1$ matrix elements
\begin{equation}\begin{split}
\langle01|H_1|11\rangle - \langle00|H_1|10\rangle =
2\frac{J}{\Delta}\frac{\delta_2}{\delta_2 + \Delta},
\end{split}
\end{equation}
which corresponds simply to the
difference of the effective Hamiltonian acting on the system with the control qubit in the
$|1\rangle$ and $|0\rangle$ states, respectively. Half this rate difference is $\mu$ which
controls the participation ratio of the CR term in the dynamics,
\begin{equation}\begin{split}
		\mu = \frac{J}{\Delta_{12}}\frac{\delta_2}{\delta_2 + \Delta_{12}}.\\
      \label{mu}\end{split}
\end{equation}
The CR effect and $\mu$ are symmetric with the exchange of target and control qubit.
However, for the remainder of this paper we will choose Q1 as the target to allow
tunning of the qubit-qubit detuning $\Delta$ and choose Q2 as the control qubit.

\section{III. Experimental Setup}
\label{sec:exp}

\begin{figure}
	\includegraphics[width=0.48\textwidth]{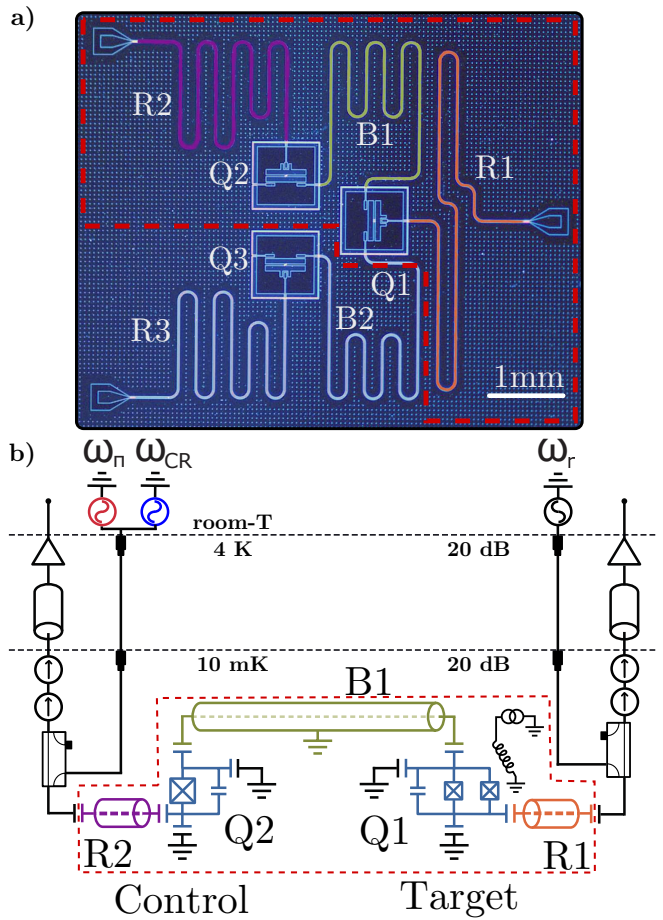}
	\caption{(color online) (a) False color optical micrograph of the device with the qubits and
resonators used in the experiment outlined in red.  (b) Circuit schematic.}
	\label{fig:schem}
\end{figure}

Our device is similar to that of Ref.~\cite{Chow2014} and consists of five
superconducting coplanar waveguide
(CPW) resonators capacitively coupled to three transmon qubits~\cite{Koch2007,Schreier2008}.
In our experiment, we use a subsection of the chip containing the central qubit and one outer
qubit, plus their associated resonators (Fig.~\ref{fig:schem}). The ground plane resonators
and qubit capacitors are fabricated from Nb on high resistivity Si. The Josephson junctions
are made with a standard shadow-evaporated Al-AlOx-Al process in a separate lithography step.
The center qubit Q1 is coupled to the two outer qubits, Q2 and Q3, via dedicated coupling
resonators, B1 and B2. Each qubit is coupled to a readout resonator, R1, R2, and R3,
which are also used for individual microwave drive. For this chip, the readout
resonators had characteristic frequency: $\omega_{\mathrm{R1}}/2\pi=6.5882\,{\rm GHz}$,
$\omega_{\mathrm{R2}}/2\pi=6.6905\,{\rm GHz}$ and line width $\kappa_{\mathrm{R1}}/2\pi=0.398\,{\rm MHz}$,
$\kappa_{\mathrm{R2}}/2\pi=0.443\,{\rm MHz}$ respectively when Q1 is tunned to at its highest transition
frequency. The bus resonator B1 was
designed to have a resonance at $\omega_{\mathrm{B1}}/2\pi=8\,{\rm GHz}$ but was not measured. Q2 is a
single junction transmon and thus has a fixed transition frequency between the ground and
excited state, $\omega_{ge}$. Q1 is a split-junction transmon and contains two junctions
arranged in a Superconducting QUantum Interference Device (SQUID) geometry to allow for
tuning of the transition frequency with an external magnetic flux~\cite{Koch2007}. The unused
resonator and qubit, R3 and Q3, were measured to have resonance frequencies of
$\omega_{\mathrm{R3}}/2\pi=6.7189\,{\rm GHz}$ and $\omega_{\mathrm{Q3}}/2\pi=4.3435\,{\rm GHz}$ respectively.
To measure the state of each qubit, we employ an autodyne measurement~\cite{Ryan2015} of the
resonator reflection where the state of the qubit is imprinted on the
dispersive shift of its corresponding readout resonances. Drive and measurement tones are shaped using
custom arbitrary waveform generators (APS) available from Raytheon BBN~\cite{BBNAPS}.
Microwave signals are sent to the
CPW readout resonators through directional couplers.  The reflected signals pass through a series
of microwave isolators before being amplified by a HEMT at the 3K stage. The signals are then
further amplified at room temperature before being down converted with a
doubly-balanced mixer and digitized by an Alazar
ATS9870 data acquisition card.

\begin{figure}
	\includegraphics[width=0.48\textwidth]{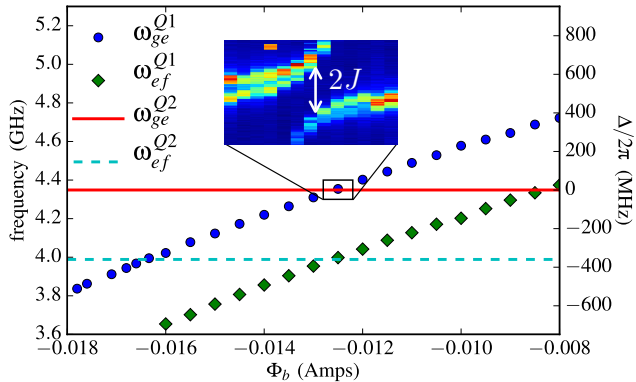}
	\caption{Spectroscopy showing the flux-modulation of the tunable qubit (Q1) above
and below the fixed-frequency qubit (Q2).  The inset shows the avoided crossing between
the two qubits.  From this we extract a $J/2\pi$ coupling of $1.08\,{\rm
	MHz}$ between the qubits.  Horizontal lines denote the 0-1 transition frequency $\omega_{ge}$ (solid red) and the 1-2 transition frequency $\omega_{ef}$ (dashed blue) of Q2.}
	\label{fig:spec}
\end{figure}

We perform spectroscopy on R1
and R2 to identify the qubit transition frequencies (Fig.~\ref{fig:spec}). For Q2, we observe
$\omega_{ge}^{Q2}/2\pi=4.349\,{\rm GHz}$ with an anharmonicity $\delta_{2}/2\pi=(
\omega_{ge}^{Q2}-\omega_{ef}^{Q2})/2\pi= -360\,{\rm MHz}$. We then step through the applied
magnetic flux $\Phi_{Q1}$ to follow the variation in transition frequency for split-junction qubit Q1.
The qubit has an upper, flux-insensitive sweet spot at $\omega_{ge}^{Q1}=5.786\,{\rm GHz}$
and an anharmonicity $\delta_{1}/2\pi = -347\,{\rm MHz}$. The modulation allows
$\omega_{ge}^{Q1}$ to be adjusted to various detunings around $\omega_{ge}^{Q2}$. We extract
a qubit-qubit coupling $J/2\pi = 1.08\,{\rm MHz} \pm 0.1\,{\rm MHz}$ by tuning the qubits on
resonance with each other and observing their anticrossing in spectroscopy \cite{Majer2007}
(Fig.~\ref{fig:spec}).

An energy relaxation time
$T_1 = 57\,\mu{\rm s}$ and dephasing time $T_2^* =
7.8\,\mu{\rm s}$ were observed for Q2. The short dephasing time is likely due to the large charging
energy $E_C$ and the associated enhanced sensitivity to charge noise~\cite{Koch2007}. A $\sim$400
kHz charge splitting was observed in spectroscopy for Q2. For Q1, we measure $T_1 = 50\,
\mu{\rm s}$ and $T_2^* = 2.8\,\mu{\rm s}$ on average in the region where our experiments were
performed. The reduced dephasing time of Q1 can be attributed to its sensitivity to magnetic
flux noise at the bias points far from the flux-insensitive sweetspot where our
experiment was conducted.

\section{IV. Measurements of Cross Resonance}
\label{sec:meas}

\begin{figure}
	\includegraphics[width=0.48\textwidth]{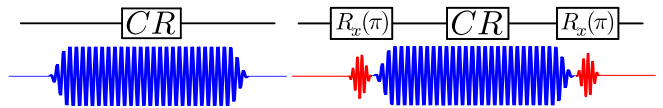}
	\caption{(color online) Pulse schematic showing the CR drive signals with the
	control in the ground (left) and excited (right) state.}
	\label{fig:pulse_schem}
\end{figure}
\begin{figure}
	\includegraphics[width=0.48\textwidth]{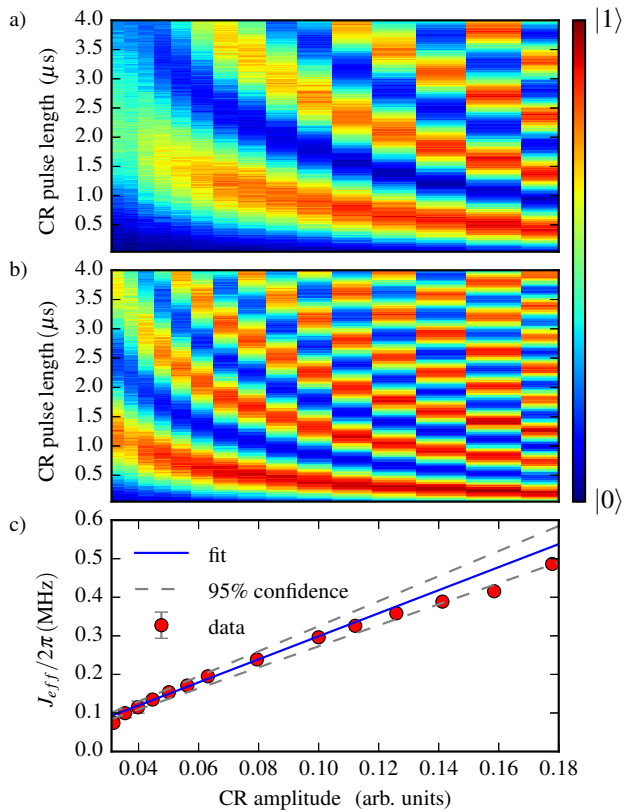}
	\caption{(color online) Measurements of Rabi oscillations of the target qubit
during the CR
	drive for both states of the control qubit at $\Delta = -78\,{\rm MHz}$.  (a) Density plot showing the
probability of finding the target qubit Q1 in the excited state as a function of CR
pulse amplitude and duration.  (b) Density plot showing the
probability of finding the target qubit Q1 in the excited state as a function of CR
pulse amplitude and duration but with a $\pi$ pulse applied before and after the CR
pulse.  (c) Plot of $J_{\mathrm{eff}}/2\pi$ vs CR pulse amplitude
  from figures (b) and (a) as described in the text with a fit to the linear regime.
The dashed lines correspond to 95\% confidence intervals on the fit for the given
data.}
	\label{fig:CR-density}
\end{figure}

Following our spectroscopic map of the two qubits, we move the tunable
qubit to a specific detuning and apply a microwave pulse of frequency
$\omega_{ge}^{Q1}(\Phi_b)/2\pi$ to the drive line of Q2. To see the CR effect, we perform a
Rabi-style measurement where we scan through the pulse width for the microwave signal at
$\omega_{ge}^{Q1}(\Phi_b)/2\pi$ applied to the drive line coupled to Q2 with no pulses
applied at $\omega_{ge}^{Q2}/2\pi$(Fig.~\ref{fig:pulse_schem}).  Thus, Q2 ideally remains in its ground
state during the sequence. We then repeat this measurement, but with a $\pi$-pulse applied to
Q2 with the microwave generator tuned to $\omega_{ge}^{Q2}/2\pi$ before and after the CR
pulse (Fig.~\ref{fig:pulse_schem}). This drives Q2 into its first excited state before the
application of the CR pulse and returns Q2 to the ground state after the CR pulse.
Immediately following this pulse sequence, the state of Q1 is readout through its
readout
resonator. Rabi oscillations of Q1 are plotted in (Fig.~\ref{fig:CR-density}) verse
drive power. The sweep of CR drive power allows us to extract the quantity $\mu$ which
will be detailed in the next section.  We then repeat
this process while stepping through applied magnetic flux to vary the qubit-qubit detuning
$\Delta/2\pi=( \omega_{ge}^{Q1}(\Phi_b) - \omega_{ge}^{Q2})/2\pi$.

\section{V. Analysis}
\label{sec:analysis}

\begin{figure}
	\includegraphics[width=0.52\textwidth]{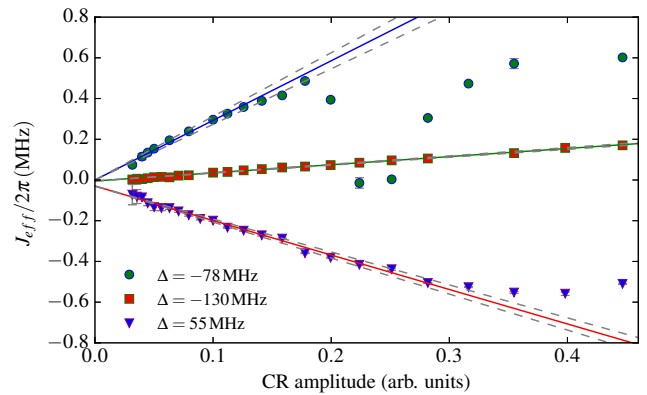}
	\caption{(color online) Plots of $J_{\mathrm{eff}}$ vs. amplitude for three select detunings with fast,
	slow and negative CR. Error bars for the points indicate uncertainty in the frequency
	difference while the dashed grey lines indicate 95\% confidence intervals for the slope
	in the linear region.}\label{fig:J_eff}
\end{figure}

At each detuning we analyze the CR data by fitting damped sinusoids to the oscillations for
each drive amplitude via the least-squares method. Subtracting the extracted oscillation
frequencies from the traces with and without $\pi$-pulses on the control qubit, we obtain
twice the CR interaction strength for each drive amplitude.
\begin{equation}\begin{split}
	J_{\mathrm{eff}}/2\pi = \frac{f_{Rabi}^{\pi}-f_{Rabi}^{}}{2}
      \label{Jeff}\end{split}
\end{equation}
The frequency difference is twice $J_{\mathrm{eff}}$ since the target qubit rotates in
the opposite direction and at equal
rates dependent on the state of the control qubit. At low powers the CR
response is linear in drive amplitude. The frequency difference at low drive amplitude is
plotted in (Fig~\ref{fig:CR-density}{\red c}) for a detuning of $-78\,{\rm MHz}$. This
linear response is plotted
for other values of $\Delta$ in (Fig.~\ref{fig:J_eff}) where we plot $J_{\mathrm{eff}}$ at three different
detunings, one corresponding to a fast CR rate, another to a slow CR rate and a third to a CR
rate with the opposite sign. In general, we
observe a linear increase in $J_{\mathrm{eff}}$ for small amplitudes, followed by a saturation at
larger amplitudes, similar to the behavior in Ref.~\cite{Chow2011}. However, the slope at
small amplitudes and the magnitude of the saturation value of $J_{\mathrm{eff}}$ depend strongly on
$\Delta$.

\begin{figure}
	\includegraphics[width=0.48\textwidth]{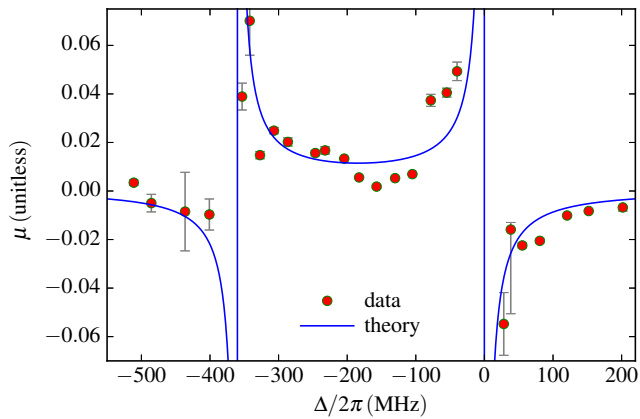}
	\caption{(color online) Plot of the extracted $\mu$ values.  The error bars are
  the confidence intervals from (Fig.~\ref{fig:J_eff}). Theoretical curve for $\mu$
vs. detuning from Eq.\eqref{mu} with data scaled by a factor computed as described in text.}
	\label{fig:mu_vs_detuning}
\end{figure}

In (Fig~\ref{fig:J_eff}) we clearly see an drop in $J_{\mathrm{eff}}$ to a small value before
recovering and behaving more erratically.  The dip in $J_{\mathrm{eff}}$ was observed
in all traces of $J_{\mathrm{eff}}$ (Fig.~\ref{fig:J_eff}) that reached saturation.
Similar behavior was observed in numerical simulations but was very sensitive to the
choice of parameters making direct comparison impossible.  Further theory of the CR effect at high drive power has been undertaken~\cite{Tripathi2019} which might explain this behavior with leakage being a likely cause. 
The slope of a linear fit to $J_{\mathrm{eff}}$ at small amplitudes yields the CR parameter $\mu$.
Extracting $\mu$ at each $\Delta$ we obtain a plot of the CR rate vs. qubit-qubit
detuning (Fig.~\ref{fig:mu_vs_detuning}). The error bars
correspond to the 95\% confidence intervals on the linear slope in the low amplitude regime which
are the dashed lines in (Fig.~\ref{fig:J_eff}).

\begin{figure}
	\includegraphics[width=0.48\textwidth]{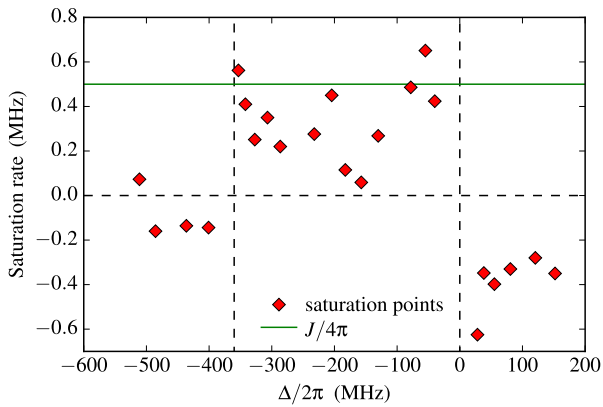}
	\caption{(color online) Saturation rate vs. detuning extracted from the $J_{\mathrm{eff}}$ plots.
  Dashed lines show the $\Delta = 0\,{\rm MHz}$ and $\Delta = 360\,{\rm MHz}$ points
  as well as a line to indicate $J_{\mathrm{eff}} = 0$.  The green line denotes the bare coupling $J$ between the qubits when tuned in resonance with each other.}
	\label{fig:saturation}
\end{figure}

To relate these slopes to Eq.~\eqref{mu}, we need to normalize them by a quantity
capturing the drive susceptibility of the control qubit.
To this end, the data points in (Fig.~\ref{fig:saturation}) are scaled by a
single best fit parameter to the theory curve. For this data a value of 75.5 MHz/amp was found.
The values of the theory curve are well known
since the chip parameters needed to calculate it are easily measured experimental quantities,
in this case the always-on $J$ coupling, control qubit anharmonicity and the
qubit-qubit detuning. An
analysis of the drive line and electronics predicted a value of $\sim$98
MHz/amp. This is 20\% from the estimated value showing the fitted parameter it to be reasonable.

In addition to extracting $\mu(\Delta)$ from the $J_{\mathrm{eff}}$ plots, we also measure the
CR saturation rate for each $\Delta$ by fitting a line with zero slope to the
asymptotic level of $J_{\mathrm{eff}}$ for large amplitudes. This corresponds to the
quickest
the CR interaction can be driven in practical experiments for a given chip and
detuning.  Figure \ref{fig:saturation} shows the
saturation rate vs. $\Delta$.  Regions near the control qubit's 0-1 transition and its
anharmonicity tend to saturate at the highest rates.  This is consistent with the measured
$\mu$ values as the two should be roughly correlated.  In contrast to $\mu$ there is no analytic
expression for the saturation rate, making further analysis difficult though it shouldn't be faster 
than the bare couping $J$ between qubits.
\section {VI. Conclusions}
In conclusion, we have explored the CR effect as a function of qubit-qubit detuning. This
work represents the first systematic study of CR vs. $\Delta$ on a single chip. We find good
agreement between the experimental data and a model accounting for the higher energy levels of
the transmon. This work will help guide future chip design by highlighting regions where the
CR rate can be increased by controlling the relative detuning between qubits in a mulit-qubit
system. With increasing qubit density and chip complexity, a clear understanding of this
aspect of frequency space will only become more important.
\section{Acknowledgments}
We thank M. B. Rothwell and G. A. Keefe for fabricating devices.
This research was funded by the Office of the Director of National Intelligence (ODNI),
Intelligence Advanced Research Projects Activity (IARPA), through the Army Research Office
under grant No. W911NF-10-1-0324. All statements of fact, opinion or conclusions contained
herein are those of the authors and should not be construed as representing the official
views or policies of IARPA, the ODNI, or the U.S. Government. Some of the preliminary device
fabrication involved the use of the Cornell NanoScale Facility, a member of the National
Nanotechnology Infrastructure Network, which is supported by the National Science Foundation
(Grant ECS-0335765).
\bibliography{tunable-cr}

\begin{thebibliography}{20}%
\makeatletter
\providecommand \@ifxundefined [1]{%
 \@ifx{#1\undefined}
}%
\providecommand \@ifnum [1]{%
 \ifnum #1\expandafter \@firstoftwo
 \else \expandafter \@secondoftwo
 \fi
}%
\providecommand \@ifx [1]{%
 \ifx #1\expandafter \@firstoftwo
 \else \expandafter \@secondoftwo
 \fi
}%
\providecommand \natexlab [1]{#1}%
\providecommand \enquote  [1]{``#1''}%
\providecommand \bibnamefont  [1]{#1}%
\providecommand \bibfnamefont [1]{#1}%
\providecommand \citenamefont [1]{#1}%
\providecommand \href@noop [0]{\@secondoftwo}%
\providecommand \href [0]{\begingroup \@sanitize@url \@href}%
\providecommand \@href[1]{\@@startlink{#1}\@@href}%
\providecommand \@@href[1]{\endgroup#1\@@endlink}%
\providecommand \@sanitize@url [0]{\catcode `\\12\catcode `\$12\catcode
  `\&12\catcode `\#12\catcode `\^12\catcode `\_12\catcode `\%12\relax}%
\providecommand \@@startlink[1]{}%
\providecommand \@@endlink[0]{}%
\providecommand \url  [0]{\begingroup\@sanitize@url \@url }%
\providecommand \@url [1]{\endgroup\@href {#1}{\urlprefix }}%
\providecommand \urlprefix  [0]{URL }%
\providecommand \Eprint [0]{\href }%
\providecommand \doibase [0]{http://dx.doi.org/}%
\providecommand \selectlanguage [0]{\@gobble}%
\providecommand \bibinfo  [0]{\@secondoftwo}%
\providecommand \bibfield  [0]{\@secondoftwo}%
\providecommand \translation [1]{[#1]}%
\providecommand \BibitemOpen [0]{}%
\providecommand \bibitemStop [0]{}%
\providecommand \bibitemNoStop [0]{.\EOS\space}%
\providecommand \EOS [0]{\spacefactor3000\relax}%
\providecommand \BibitemShut  [1]{\csname bibitem#1\endcsname}%
\let\auto@bib@innerbib\@empty
\bibitem [{\citenamefont {Clarke}\ and\ \citenamefont
  {Wilhelm}(2008)}]{Clarke2008}%
  \BibitemOpen
  \bibfield  {author} {\bibinfo {author} {\bibfnamefont {J.}~\bibnamefont
  {Clarke}}\ and\ \bibinfo {author} {\bibfnamefont {F.~K.}\ \bibnamefont
  {Wilhelm}},\ }\href {http://dx.doi.org/10.1038/nature07128} {\bibfield
  {journal} {\bibinfo  {journal} {Nature}\ }\textbf {\bibinfo {volume} {453}},\
  \bibinfo {pages} {1031} (\bibinfo {year} {2008})}\BibitemShut {NoStop}%
\bibitem [{\citenamefont {Devoret}\ and\ \citenamefont
  {Schoelkopf}(2013)}]{Devoret2013}%
  \BibitemOpen
  \bibfield  {author} {\bibinfo {author} {\bibfnamefont {M.~H.}\ \bibnamefont
  {Devoret}}\ and\ \bibinfo {author} {\bibfnamefont {R.~J.}\ \bibnamefont
  {Schoelkopf}},\ }\href {\doibase 10.1126/science.1231930} {\bibfield
  {journal} {\bibinfo  {journal} {Science}\ }\textbf {\bibinfo {volume}
  {339}},\ \bibinfo {pages} {1169} (\bibinfo {year} {2013})}\BibitemShut
  {NoStop}%
\bibitem [{\citenamefont {Chow}\ \emph {et~al.}(2014)\citenamefont {Chow},
  \citenamefont {Gambetta}, \citenamefont {Magesan}, \citenamefont {Abraham},
  \citenamefont {Cross}, \citenamefont {Johnson}, \citenamefont {Masluk},
  \citenamefont {Ryan}, \citenamefont {Smolin}, \citenamefont {Srinivasan},\
  and\ \citenamefont {Steffen}}]{Chow2014}%
  \BibitemOpen
  \bibfield  {author} {\bibinfo {author} {\bibfnamefont {J.~M.}\ \bibnamefont
  {Chow}}, \bibinfo {author} {\bibfnamefont {J.~M.}\ \bibnamefont {Gambetta}},
  \bibinfo {author} {\bibfnamefont {E.}~\bibnamefont {Magesan}}, \bibinfo
  {author} {\bibfnamefont {D.~W.}\ \bibnamefont {Abraham}}, \bibinfo {author}
  {\bibfnamefont {A.~W.}\ \bibnamefont {Cross}}, \bibinfo {author}
  {\bibfnamefont {B.~R.}\ \bibnamefont {Johnson}}, \bibinfo {author}
  {\bibfnamefont {N.~A.}\ \bibnamefont {Masluk}}, \bibinfo {author}
  {\bibfnamefont {C.~A.}\ \bibnamefont {Ryan}}, \bibinfo {author}
  {\bibfnamefont {J.~A.}\ \bibnamefont {Smolin}}, \bibinfo {author}
  {\bibfnamefont {S.~J.}\ \bibnamefont {Srinivasan}}, \ and\ \bibinfo {author}
  {\bibfnamefont {M.}~\bibnamefont {Steffen}},\ }\href
  {http://dx.doi.org/10.1038/ncomms5015} {\bibfield  {journal} {\bibinfo
  {journal} {Nat Commun}\ }\textbf {\bibinfo {volume} {5:4015}} (\bibinfo
  {year} {2014})}\BibitemShut {NoStop}%
\bibitem [{\citenamefont {Barends}\ \emph {et~al.}(2014)\citenamefont
  {Barends}, \citenamefont {Kelly}, \citenamefont {Megrant}, \citenamefont
  {Veitia}, \citenamefont {Sank}, \citenamefont {Jeffrey}, \citenamefont
  {White}, \citenamefont {Mutus}, \citenamefont {Fowler}, \citenamefont
  {Campbell}, \citenamefont {Chen}, \citenamefont {Chen}, \citenamefont
  {Chiaro}, \citenamefont {Dunsworth}, \citenamefont {Neill}, \citenamefont
  {O'Malley}, \citenamefont {Roushan}, \citenamefont {Vainsencher},
  \citenamefont {Wenner}, \citenamefont {Korotkov}, \citenamefont {Cleland},\
  and\ \citenamefont {Martinis}}]{Barends2014}%
  \BibitemOpen
  \bibfield  {author} {\bibinfo {author} {\bibfnamefont {R.}~\bibnamefont
  {Barends}}, \bibinfo {author} {\bibfnamefont {J.}~\bibnamefont {Kelly}},
  \bibinfo {author} {\bibfnamefont {A.}~\bibnamefont {Megrant}}, \bibinfo
  {author} {\bibfnamefont {A.}~\bibnamefont {Veitia}}, \bibinfo {author}
  {\bibfnamefont {D.}~\bibnamefont {Sank}}, \bibinfo {author} {\bibfnamefont
  {E.}~\bibnamefont {Jeffrey}}, \bibinfo {author} {\bibfnamefont {T.~C.}\
  \bibnamefont {White}}, \bibinfo {author} {\bibfnamefont {J.}~\bibnamefont
  {Mutus}}, \bibinfo {author} {\bibfnamefont {A.~G.}\ \bibnamefont {Fowler}},
  \bibinfo {author} {\bibfnamefont {B.}~\bibnamefont {Campbell}}, \bibinfo
  {author} {\bibfnamefont {Y.}~\bibnamefont {Chen}}, \bibinfo {author}
  {\bibfnamefont {Z.}~\bibnamefont {Chen}}, \bibinfo {author} {\bibfnamefont
  {B.}~\bibnamefont {Chiaro}}, \bibinfo {author} {\bibfnamefont
  {A.}~\bibnamefont {Dunsworth}}, \bibinfo {author} {\bibfnamefont
  {C.}~\bibnamefont {Neill}}, \bibinfo {author} {\bibfnamefont
  {P.}~\bibnamefont {O'Malley}}, \bibinfo {author} {\bibfnamefont
  {P.}~\bibnamefont {Roushan}}, \bibinfo {author} {\bibfnamefont
  {A.}~\bibnamefont {Vainsencher}}, \bibinfo {author} {\bibfnamefont
  {J.}~\bibnamefont {Wenner}}, \bibinfo {author} {\bibfnamefont {A.~N.}\
  \bibnamefont {Korotkov}}, \bibinfo {author} {\bibfnamefont {A.~N.}\
  \bibnamefont {Cleland}}, \ and\ \bibinfo {author} {\bibfnamefont {J.~M.}\
  \bibnamefont {Martinis}},\ }\href {http://dx.doi.org/10.1038/nature13171}
  {\bibfield  {journal} {\bibinfo  {journal} {Nature}\ }\textbf {\bibinfo
  {volume} {508}},\ \bibinfo {pages} {500} (\bibinfo {year}
  {2014})}\BibitemShut {NoStop}%
\bibitem [{\citenamefont {C\'{o}rcoles}\ \emph {et~al.}(2015)\citenamefont
  {C\'{o}rcoles}, \citenamefont {Magesan}, \citenamefont {Srinivasan},
  \citenamefont {Cross}, \citenamefont {Steffen}, \citenamefont {Gambetta},\
  and\ \citenamefont {Chow}}]{Corcoles2015}%
  \BibitemOpen
  \bibfield  {author} {\bibinfo {author} {\bibfnamefont {A.}~\bibnamefont
  {C\'{o}rcoles}}, \bibinfo {author} {\bibfnamefont {E.}~\bibnamefont
  {Magesan}}, \bibinfo {author} {\bibfnamefont {S.~J.}\ \bibnamefont
  {Srinivasan}}, \bibinfo {author} {\bibfnamefont {A.~W.}\ \bibnamefont
  {Cross}}, \bibinfo {author} {\bibfnamefont {M.}~\bibnamefont {Steffen}},
  \bibinfo {author} {\bibfnamefont {J.~M.}\ \bibnamefont {Gambetta}}, \ and\
  \bibinfo {author} {\bibfnamefont {J.~M.}\ \bibnamefont {Chow}},\ }\href
  {http://dx.doi.org/10.1038/ncomms7979} {\bibfield  {journal} {\bibinfo
  {journal} {Nat Commun}\ }\textbf {\bibinfo {volume} {6}} (\bibinfo {year}
  {2015})}\BibitemShut {NoStop}%
\bibitem [{\citenamefont {Fowler}\ \emph {et~al.}(2012)\citenamefont {Fowler},
  \citenamefont {Mariantoni}, \citenamefont {Martinis},\ and\ \citenamefont
  {Cleland}}]{Fowler2012}%
  \BibitemOpen
  \bibfield  {author} {\bibinfo {author} {\bibfnamefont {A.~G.}\ \bibnamefont
  {Fowler}}, \bibinfo {author} {\bibfnamefont {M.}~\bibnamefont {Mariantoni}},
  \bibinfo {author} {\bibfnamefont {J.~M.}\ \bibnamefont {Martinis}}, \ and\
  \bibinfo {author} {\bibfnamefont {A.~N.}\ \bibnamefont {Cleland}},\ }\href
  {\doibase 10.1103/PhysRevA.86.032324} {\bibfield  {journal} {\bibinfo
  {journal} {Phys. Rev. A}\ }\textbf {\bibinfo {volume} {86}},\ \bibinfo
  {pages} {032324} (\bibinfo {year} {2012})}\BibitemShut {NoStop}%
\bibitem [{\citenamefont {Rigetti}\ and\ \citenamefont
  {Devoret}(2010)}]{Rigetti2010}%
  \BibitemOpen
  \bibfield  {author} {\bibinfo {author} {\bibfnamefont {C.}~\bibnamefont
  {Rigetti}}\ and\ \bibinfo {author} {\bibfnamefont {M.}~\bibnamefont
  {Devoret}},\ }\href {\doibase 10.1103/PhysRevB.81.134507} {\bibfield
  {journal} {\bibinfo  {journal} {Phys. Rev. B}\ }\textbf {\bibinfo {volume}
  {81}},\ \bibinfo {pages} {134507} (\bibinfo {year} {2010})}\BibitemShut
  {NoStop}%
\bibitem [{\citenamefont {Paraoanu}(2006)}]{Paraoanu2006}%
  \BibitemOpen
  \bibfield  {author} {\bibinfo {author} {\bibfnamefont {G.~S.}\ \bibnamefont
  {Paraoanu}},\ }\href {\doibase 10.1103/PhysRevB.74.140504} {\bibfield
  {journal} {\bibinfo  {journal} {Phys. Rev. B}\ }\textbf {\bibinfo {volume}
  {74}},\ \bibinfo {pages} {140504} (\bibinfo {year} {2006})}\BibitemShut
  {NoStop}%
\bibitem [{\citenamefont {Chow}\ \emph {et~al.}(2011)\citenamefont {Chow},
  \citenamefont {C\'orcoles}, \citenamefont {Gambetta}, \citenamefont
  {Rigetti}, \citenamefont {Johnson}, \citenamefont {Smolin}, \citenamefont
  {Rozen}, \citenamefont {Keefe}, \citenamefont {Rothwell}, \citenamefont
  {Ketchen},\ and\ \citenamefont {Steffen}}]{Chow2011}%
  \BibitemOpen
  \bibfield  {author} {\bibinfo {author} {\bibfnamefont {J.~M.}\ \bibnamefont
  {Chow}}, \bibinfo {author} {\bibfnamefont {A.~D.}\ \bibnamefont
  {C\'orcoles}}, \bibinfo {author} {\bibfnamefont {J.~M.}\ \bibnamefont
  {Gambetta}}, \bibinfo {author} {\bibfnamefont {C.}~\bibnamefont {Rigetti}},
  \bibinfo {author} {\bibfnamefont {B.~R.}\ \bibnamefont {Johnson}}, \bibinfo
  {author} {\bibfnamefont {J.~A.}\ \bibnamefont {Smolin}}, \bibinfo {author}
  {\bibfnamefont {J.~R.}\ \bibnamefont {Rozen}}, \bibinfo {author}
  {\bibfnamefont {G.~A.}\ \bibnamefont {Keefe}}, \bibinfo {author}
  {\bibfnamefont {M.~B.}\ \bibnamefont {Rothwell}}, \bibinfo {author}
  {\bibfnamefont {M.~B.}\ \bibnamefont {Ketchen}}, \ and\ \bibinfo {author}
  {\bibfnamefont {M.}~\bibnamefont {Steffen}},\ }\href {\doibase
  10.1103/PhysRevLett.107.080502} {\bibfield  {journal} {\bibinfo  {journal}
  {Phys. Rev. Lett.}\ }\textbf {\bibinfo {volume} {107}},\ \bibinfo {pages}
  {080502} (\bibinfo {year} {2011})}\BibitemShut {NoStop}%
\bibitem [{\citenamefont {C\'orcoles}\ \emph {et~al.}(2013)\citenamefont
  {C\'orcoles}, \citenamefont {Gambetta}, \citenamefont {Chow}, \citenamefont
  {Smolin}, \citenamefont {Ware}, \citenamefont {Strand}, \citenamefont
  {Plourde},\ and\ \citenamefont {Steffen}}]{Corcoles2013}%
  \BibitemOpen
  \bibfield  {author} {\bibinfo {author} {\bibfnamefont {A.~D.}\ \bibnamefont
  {C\'orcoles}}, \bibinfo {author} {\bibfnamefont {J.~M.}\ \bibnamefont
  {Gambetta}}, \bibinfo {author} {\bibfnamefont {J.~M.}\ \bibnamefont {Chow}},
  \bibinfo {author} {\bibfnamefont {J.~A.}\ \bibnamefont {Smolin}}, \bibinfo
  {author} {\bibfnamefont {M.}~\bibnamefont {Ware}}, \bibinfo {author}
  {\bibfnamefont {J.}~\bibnamefont {Strand}}, \bibinfo {author} {\bibfnamefont
  {B.~L.~T.}\ \bibnamefont {Plourde}}, \ and\ \bibinfo {author} {\bibfnamefont
  {M.}~\bibnamefont {Steffen}},\ }\href {\doibase 10.1103/PhysRevA.87.030301}
  {\bibfield  {journal} {\bibinfo  {journal} {Phys. Rev. A}\ }\textbf {\bibinfo
  {volume} {87}},\ \bibinfo {pages} {030301} (\bibinfo {year}
  {2013})}\BibitemShut {NoStop}%
\bibitem [{\citenamefont {Majer}\ \emph {et~al.}(2007)\citenamefont {Majer},
  \citenamefont {Chow}, \citenamefont {Gambetta}, \citenamefont {Koch},
  \citenamefont {Johnson}, \citenamefont {Schreier}, \citenamefont {Frunzio},
  \citenamefont {Schuster}, \citenamefont {Houck}, \citenamefont {Wallraff},
  \citenamefont {Blais}, \citenamefont {Devoret}, \citenamefont {Girvin},\ and\
  \citenamefont {Schoelkopf}}]{Majer2007}%
  \BibitemOpen
  \bibfield  {author} {\bibinfo {author} {\bibfnamefont {J.}~\bibnamefont
  {Majer}}, \bibinfo {author} {\bibfnamefont {J.~M.}\ \bibnamefont {Chow}},
  \bibinfo {author} {\bibfnamefont {J.~M.}\ \bibnamefont {Gambetta}}, \bibinfo
  {author} {\bibfnamefont {J.}~\bibnamefont {Koch}}, \bibinfo {author}
  {\bibfnamefont {B.~R.}\ \bibnamefont {Johnson}}, \bibinfo {author}
  {\bibfnamefont {J.~A.}\ \bibnamefont {Schreier}}, \bibinfo {author}
  {\bibfnamefont {L.}~\bibnamefont {Frunzio}}, \bibinfo {author} {\bibfnamefont
  {D.~I.}\ \bibnamefont {Schuster}}, \bibinfo {author} {\bibfnamefont {A.~A.}\
  \bibnamefont {Houck}}, \bibinfo {author} {\bibfnamefont {A.}~\bibnamefont
  {Wallraff}}, \bibinfo {author} {\bibfnamefont {A.}~\bibnamefont {Blais}},
  \bibinfo {author} {\bibfnamefont {M.~H.}\ \bibnamefont {Devoret}}, \bibinfo
  {author} {\bibfnamefont {S.~M.}\ \bibnamefont {Girvin}}, \ and\ \bibinfo
  {author} {\bibfnamefont {R.~J.}\ \bibnamefont {Schoelkopf}},\ }\href
  {\doibase 10.1038/nature06184} {\bibfield  {journal} {\bibinfo  {journal}
  {Nature}\ }\textbf {\bibinfo {volume} {449}},\ \bibinfo {pages} {443}
  (\bibinfo {year} {2007})}\BibitemShut {NoStop}%
\bibitem [{\citenamefont {Gambetta}(2013)}]{Gambetta2013}%
  \BibitemOpen
  \bibfield  {author} {\bibinfo {author} {\bibfnamefont {J.}~\bibnamefont
  {Gambetta}},\ }\href@noop {} {\bibfield  {journal} {\bibinfo  {journal} {IFF
  Spring School 2013 Quantum Information Processing Lecture Notes}\ }\textbf
  {\bibinfo {volume} {339}},\ \bibinfo {pages} {Chap. B4} (\bibinfo {year}
  {2013})}\BibitemShut {NoStop}%
\bibitem [{\citenamefont {Magesan}\ and\ \citenamefont
  {Gambetta}(2018)}]{Magesan2019}%
  \BibitemOpen
  \bibfield  {author} {\bibinfo {author} {\bibfnamefont {E.}~\bibnamefont
  {Magesan}}\ and\ \bibinfo {author} {\bibfnamefont {J.~M.}\ \bibnamefont
  {Gambetta}},\ }\href@noop {} {\  (\bibinfo {year} {2018})},\ \Eprint
  {http://arxiv.org/abs/arXiv:1804.04073} {arXiv:1804.04073} \BibitemShut
  {NoStop}%
\bibitem [{\citenamefont {de~Groot}\ \emph {et~al.}(2012)\citenamefont
  {de~Groot}, \citenamefont {Ashhab}, \citenamefont {Lupaşcu}, \citenamefont
  {DiCarlo}, \citenamefont {Nori}, \citenamefont {Harmans},\ and\ \citenamefont
  {Mooij}}]{deGroot2012}%
  \BibitemOpen
  \bibfield  {author} {\bibinfo {author} {\bibfnamefont {P.~C.}\ \bibnamefont
  {de~Groot}}, \bibinfo {author} {\bibfnamefont {S.}~\bibnamefont {Ashhab}},
  \bibinfo {author} {\bibfnamefont {A.}~\bibnamefont {Lupaşcu}}, \bibinfo
  {author} {\bibfnamefont {L.}~\bibnamefont {DiCarlo}}, \bibinfo {author}
  {\bibfnamefont {F.}~\bibnamefont {Nori}}, \bibinfo {author} {\bibfnamefont
  {C.~J. P.~M.}\ \bibnamefont {Harmans}}, \ and\ \bibinfo {author}
  {\bibfnamefont {J.~E.}\ \bibnamefont {Mooij}},\ }\href
  {http://stacks.iop.org/1367-2630/14/i=7/a=073038} {\bibfield  {journal}
  {\bibinfo  {journal} {New Journal of Physics}\ }\textbf {\bibinfo {volume}
  {14}},\ \bibinfo {pages} {073038} (\bibinfo {year} {2012})}\BibitemShut
  {NoStop}%
\bibitem [{\citenamefont {Patterson}\ \emph {et~al.}(2019)\citenamefont
  {Patterson}, \citenamefont {Rahamim}, \citenamefont {Tsunoda}, \citenamefont
  {Spring}, \citenamefont {Jebari}, \citenamefont {Ratter}, \citenamefont
  {Mergenthaler}, \citenamefont {Tancredi}, \citenamefont {Vlastakis},
  \citenamefont {Esposito},\ and\ \citenamefont {Leek}}]{Patterson2019}%
  \BibitemOpen
  \bibfield  {author} {\bibinfo {author} {\bibfnamefont {A.~D.}\ \bibnamefont
  {Patterson}}, \bibinfo {author} {\bibfnamefont {J.}~\bibnamefont {Rahamim}},
  \bibinfo {author} {\bibfnamefont {T.}~\bibnamefont {Tsunoda}}, \bibinfo
  {author} {\bibfnamefont {P.}~\bibnamefont {Spring}}, \bibinfo {author}
  {\bibfnamefont {S.}~\bibnamefont {Jebari}}, \bibinfo {author} {\bibfnamefont
  {K.}~\bibnamefont {Ratter}}, \bibinfo {author} {\bibfnamefont
  {M.}~\bibnamefont {Mergenthaler}}, \bibinfo {author} {\bibfnamefont
  {G.}~\bibnamefont {Tancredi}}, \bibinfo {author} {\bibfnamefont
  {B.}~\bibnamefont {Vlastakis}}, \bibinfo {author} {\bibfnamefont
  {M.}~\bibnamefont {Esposito}}, \ and\ \bibinfo {author} {\bibfnamefont
  {P.~J.}\ \bibnamefont {Leek}},\ }\href@noop {} {\  (\bibinfo {year}
  {2019})},\ \Eprint {http://arxiv.org/abs/arXiv:1905.05670} {arXiv:1905.05670}
  \BibitemShut {NoStop}%
\bibitem [{\citenamefont {Koch}\ \emph {et~al.}(2007)\citenamefont {Koch},
  \citenamefont {Yu}, \citenamefont {Gambetta}, \citenamefont {Houck},
  \citenamefont {Schuster}, \citenamefont {Majer}, \citenamefont {Blais},
  \citenamefont {Devoret}, \citenamefont {Girvin},\ and\ \citenamefont
  {Schoelkopf}}]{Koch2007}%
  \BibitemOpen
  \bibfield  {author} {\bibinfo {author} {\bibfnamefont {J.}~\bibnamefont
  {Koch}}, \bibinfo {author} {\bibfnamefont {T.~M.}\ \bibnamefont {Yu}},
  \bibinfo {author} {\bibfnamefont {J.}~\bibnamefont {Gambetta}}, \bibinfo
  {author} {\bibfnamefont {A.~A.}\ \bibnamefont {Houck}}, \bibinfo {author}
  {\bibfnamefont {D.~I.}\ \bibnamefont {Schuster}}, \bibinfo {author}
  {\bibfnamefont {J.}~\bibnamefont {Majer}}, \bibinfo {author} {\bibfnamefont
  {A.}~\bibnamefont {Blais}}, \bibinfo {author} {\bibfnamefont {M.~H.}\
  \bibnamefont {Devoret}}, \bibinfo {author} {\bibfnamefont {S.~M.}\
  \bibnamefont {Girvin}}, \ and\ \bibinfo {author} {\bibfnamefont {R.~J.}\
  \bibnamefont {Schoelkopf}},\ }\href {\doibase 10.1103/PhysRevA.76.042319}
  {\bibfield  {journal} {\bibinfo  {journal} {Phys. Rev. A}\ }\textbf {\bibinfo
  {volume} {76}},\ \bibinfo {pages} {042319} (\bibinfo {year}
  {2007})}\BibitemShut {NoStop}%
\bibitem [{\citenamefont {Schreier}\ \emph {et~al.}(2008)\citenamefont
  {Schreier}, \citenamefont {Houck}, \citenamefont {Koch}, \citenamefont
  {Schuster}, \citenamefont {Johnson}, \citenamefont {Chow}, \citenamefont
  {Gambetta}, \citenamefont {Majer}, \citenamefont {Frunzio}, \citenamefont
  {Devoret}, \citenamefont {Girvin},\ and\ \citenamefont
  {Schoelkopf}}]{Schreier2008}%
  \BibitemOpen
  \bibfield  {author} {\bibinfo {author} {\bibfnamefont {J.~A.}\ \bibnamefont
  {Schreier}}, \bibinfo {author} {\bibfnamefont {A.~A.}\ \bibnamefont {Houck}},
  \bibinfo {author} {\bibfnamefont {J.}~\bibnamefont {Koch}}, \bibinfo {author}
  {\bibfnamefont {D.~I.}\ \bibnamefont {Schuster}}, \bibinfo {author}
  {\bibfnamefont {B.~R.}\ \bibnamefont {Johnson}}, \bibinfo {author}
  {\bibfnamefont {J.~M.}\ \bibnamefont {Chow}}, \bibinfo {author}
  {\bibfnamefont {J.~M.}\ \bibnamefont {Gambetta}}, \bibinfo {author}
  {\bibfnamefont {J.}~\bibnamefont {Majer}}, \bibinfo {author} {\bibfnamefont
  {L.}~\bibnamefont {Frunzio}}, \bibinfo {author} {\bibfnamefont {M.~H.}\
  \bibnamefont {Devoret}}, \bibinfo {author} {\bibfnamefont {S.~M.}\
  \bibnamefont {Girvin}}, \ and\ \bibinfo {author} {\bibfnamefont {R.~J.}\
  \bibnamefont {Schoelkopf}},\ }\href {\doibase 10.1103/PhysRevB.77.180502}
  {\bibfield  {journal} {\bibinfo  {journal} {Phys. Rev. B}\ }\textbf {\bibinfo
  {volume} {77}},\ \bibinfo {pages} {180502} (\bibinfo {year}
  {2008})}\BibitemShut {NoStop}%
\bibitem [{\citenamefont {Ryan}\ \emph {et~al.}(2015)\citenamefont {Ryan},
  \citenamefont {Johnson}, \citenamefont {Gambetta}, \citenamefont {Chow},
  \citenamefont {da~Silva}, \citenamefont {Dial},\ and\ \citenamefont
  {Ohki}}]{Ryan2015}%
  \BibitemOpen
  \bibfield  {author} {\bibinfo {author} {\bibfnamefont {C.~A.}\ \bibnamefont
  {Ryan}}, \bibinfo {author} {\bibfnamefont {B.~R.}\ \bibnamefont {Johnson}},
  \bibinfo {author} {\bibfnamefont {J.~M.}\ \bibnamefont {Gambetta}}, \bibinfo
  {author} {\bibfnamefont {J.~M.}\ \bibnamefont {Chow}}, \bibinfo {author}
  {\bibfnamefont {M.~P.}\ \bibnamefont {da~Silva}}, \bibinfo {author}
  {\bibfnamefont {O.~E.}\ \bibnamefont {Dial}}, \ and\ \bibinfo {author}
  {\bibfnamefont {T.~A.}\ \bibnamefont {Ohki}},\ }\href {\doibase
  10.1103/PhysRevA.91.022118} {\bibfield  {journal} {\bibinfo  {journal} {Phys.
  Rev. A}\ }\textbf {\bibinfo {volume} {91}},\ \bibinfo {pages} {022118}
  (\bibinfo {year} {2015})}\BibitemShut {NoStop}%
\bibitem [{BBN()}]{BBNAPS}%
  \BibitemOpen
  \href@noop {} {}\bibinfo {howpublished}
  {\url{https://www.raytheon.com/capabilities/products/quantum}}\BibitemShut
  {NoStop}%
\bibitem [{\citenamefont {Tripathi}\ \emph {et~al.}(2019)\citenamefont
  {Tripathi}, \citenamefont {Khezri},\ and\ \citenamefont
  {Korotkov}}]{Tripathi2019}%
  \BibitemOpen
  \bibfield  {author} {\bibinfo {author} {\bibfnamefont {V.}~\bibnamefont
  {Tripathi}}, \bibinfo {author} {\bibfnamefont {M.}~\bibnamefont {Khezri}}, \
  and\ \bibinfo {author} {\bibfnamefont {A.~N.}\ \bibnamefont {Korotkov}},\
  }\href@noop {} {\  (\bibinfo {year} {2019})},\ \Eprint
  {http://arxiv.org/abs/arXiv:1902.09054} {arXiv:1902.09054} \BibitemShut
  {NoStop}%
\end{thebibliography}%
\end{document}